\documentclass[referee]{raa}            

\usepackage{graphicx,times}             
\usepackage{natbib}
\usepackage{amssymb,amsmath}
\usepackage{soul}      

\bibpunct{(}{)}{;}{a}{}{,}

\usepackage[pagebackref=true]{hyperref}
\usepackage{orcidlink}

\begin{document}

  \title{Prospects for probing dark matter with filamentary 21cm emission}

   \volnopage{Vol.0 (20xx) No.0, 000--000}      
   \setcounter{page}{1}          

   \author{Yizhou Liu 
      \inst{1,2} {\orcidlink{0009-0005-8855-0748}}
   \and Yingjie Jing 
        \inst{3} {\orcidlink{0000-0003-3433-8416}}
    \and Haoran Shi
        \inst{4}
    \and Huijie Hu
        \inst{5,3} {\orcidlink{0000-0002-1908-0384}}
   }

   \institute{Institute for Frontiers in Astronomy and Astrophysics, Beijing Normal University, Beijing 102206, China;\\
        \and
             School of Physics and Astronomy, Beijing Normal University, Beijing 100875, China;\\
        \and
             National Astronomical Observatories, Chinese Academy of Sciences, Beijing 100101, China;\\
        \and
             School of Accountancy, Central University of Finance and Economics, Beijing 100081, China;\\
        \and
            School of Physics and Astronomy, Anqing Normal University, Anqing 246133, China;\\
            {\it e-mail: yzliu@bnu.edu.cn}
\vs\no
   }

\abstract{The cosmic web contains most of the matter in the Universe, and its filamentary structure is closely related to the properties of dark matter. Compared with the cold dark matter (CDM) model, warm dark matter (WDM) suppresses the formation of low-mass halos and produces smoother cosmic filaments. In this work, we explore whether filamentary 21cm emission can be used to distinguish between these two dark matter models at $z=4$ and $2.5$. 
We generate intrinsic 21cm brightness temperature maps from cosmological hydrodynamical simulations and produce mock observations by convolving them with Gaussian beams of different angular resolutions. We then compare the predicted filament signals with the expected sensitivities of SKA1-Low and SKA2-Low. 
We find that the intrinsic 21cm emission clearly reflects the different filament morphologies in the CDM and WDM models, particularly at $z=4$. However, these differences rapidly disappear as the beam size increases, especially for beam sizes larger than $6\arcsec$. Although larger beams improve the observational sensitivity, they also erase the small-scale structures that encode the dark matter information. As a result, neither SKA1-Low nor the planned SKA2-Low can directly resolve individual filaments with sufficient sensitivity to distinguish between the two models. We therefore conclude that direct imaging of cosmic filaments and constraining dark matter through their morphology are unlikely to be feasible with current and planned SKA-Low facilities.
}

   \authorrunning{Liu et al.}            
   \titlerunning{21cm filament}  

   \maketitle

%
%
\section{Introduction}           
\label{sect:intro}

The current standard cosmological model, $\Lambda$CDM, successfully reproduces a wide range of observations, including the cosmic microwave background, the large-scale distribution of galaxies, and the evolution of cosmic structures \citep{Springel2006}. In this framework, matter is organized into a network of nodes, filaments, sheets, and voids, collectively known as the cosmic web \citep{Bond1996}. Dark matter provides the gravitational backbone of this structure, while baryons subsequently collapse into the underlying dark matter potential \citep{Frenk2012}. Although the existence of dark matter is well established, its particle nature remains one of the most important open questions in modern cosmology. 

The $\Lambda$CDM scenario assumes that dark matter is cold and therefore preserves primordial density fluctuations down to extremely small scales \citep{Hofmann2001, Green2004, Diemand2005, Wang2020, Liu2024, Zheng2024}. Alternative dark matter models, such as warm dark matter (WDM), possess a finite free-streaming length that suppresses the growth of low-mass structures \citep{Colin2000, Avila-Reese2001, Bode2001, Schneider2012, Lovell2014, Ludlow2016, Bose2017, Lovell2017, Khimey2021, Paduroiu2022}. As a result, WDM predicts fewer low-mass halos and a smoother
matter distribution on small scales than cold dark matter (CDM). These differences are expected to extend beyond individual halos and influence the morphology of the cosmic web, particularly the filamentary structures at high redshifts \citep{Gao2007, Gao2015, Mocz2019}.

Several observational probes have been developed to constrain the nature of dark matter. The Ly$\alpha$ forest provides stringent limits by measuring the small-scale matter power spectrum of the intergalactic medium \citep{Hernquist1996, Viel2004, Viel2005, Boyarsky2009, Viel2013, Baur2016, Irsic2017, Garzilli2017, Garzilli2019, Garzilli2021, Villasenor2023, Irsic2024}, while gravitational lensing offers complementary constraints through the abundance of dark matter subhalos \citep{Metcalf2001, Miranda2007, Zackrisson2010, Vegetti2012, Hezaveh2016_b, Hezaveh2016_a, Li2016, Minor2017, Gilman2020}. Another promising approach is to probe the diffuse gas residing in cosmic filaments with its emission line, such as Ly$\alpha$ emission \citep{Elias2020, Witstok2021, Byrohl2023, Liu2025, Liu2026}. Hydrodynamical simulations have shown that filamentary HI distributions are sensitive to the underlying dark matter model, with WDM producing smoother filaments than CDM \citep{Gao2015}. Since the 21cm hyperfine transition directly traces neutral hydrogen, it provides a natural tracer of these diffuse structures and therefore offers a potential means of probing dark matter through the morphology of the cosmic web.

Whether such filamentary 21cm emission can be detected with sufficient fidelity to distinguish different dark matter models, however, remains unclear. Unlike statistical measurements based on the 21cm power spectrum \citep{Carucci2015, Bauer2021}, resolving individual filaments requires both higher angular resolution and higher sensitivity. While a finer synthesized beam preserves the intrinsic filament morphology, it also substantially reduces the observational sensitivity. The interplay between these two requirements has not yet been systematically quantified.
 
In this work, we investigate the observational feasibility of using filamentary 21cm emission to distinguish between CDM and WDM. We generate intrinsic HI column density and integrated 21cm brightness temperature maps from cosmological zoom-in hydrodynamical simulations of a Milky Way-sized halo evolved in both CDM and WDM cosmologies. Realistic mock observations are then produced by convolving the intrinsic emission with Gaussian beams of different angular resolutions, and the resulting signals are compared with the expected sensitivities of SKA1-Low and SKA2-Low \citep{Braun2019}. Our goal is to determine whether the characteristic filament morphology predicted by different dark matter models can be directly observed with current and planned radio facilities.

The paper is organized as follows. Section~\ref{sect:meth} describes the simulations, the 21cm brightness temperature modelling, and the generation of mock observations. Section~\ref{sect:res} presents the main results, including the intrinsic filament properties, the effects of beam smearing, and the detectability with SKA-Low. Finally, Section~\ref{sect:discussion} discusses and summarizes our findings.

\section{Methodology}
\label{sect:meth}

\subsection{Simulation}

We employ two cosmological zoom-in hydrodynamical simulations of the same Milky Way-sized halo \citep{Gao2015}, performed with {\small GADGET-3} \citep{Springel2005}. One simulation assumes the CDM model, while the other adopts a WDM cosmology. The two runs share identical initial random phases and cosmological parameters, ensuring a direct comparison between CDM and WDM. The simulations include cooling and photoheating by a spatially uniform UV/X-ray background from \citet{Haardt2001}, and a simplified prescription for star formation. Gas particles are converted into stars once both a hydrogen number density threshold $n_{\mathrm{H}}>0.1\,\mathrm{cm^{-3}}$ and an overdensity criterion $\rho/\bar \rho > 2000$ are satisfied. Stellar and AGN feedback are not included and reionization is imposed at $z=6$ in simulations.

The mass resolution is $5.16\times10^{4}\ h^{-1}\mathrm{M_{\odot}}$ for gas and $2.35\times10^{5}\ h^{-1}\mathrm{M_{\odot}}$ for dark matter. The co-moving softening length is $0.5\,h^{-1}\mathrm{kpc}$. The target system is the 'halo A' re-simulation from the Aquarius Project \citep{Springel2008}. In the WDM model, the initial linear matter power spectrum is obtained by suppressing the CDM power below the free-streaming scale, corresponding to a thermal relic particle with a mass of $1.5\,\mathrm{keV}$. While this value is below the current Ly$\alpha$ forest constraints \citep{Villasenor2023, Irsic2024}, it is adopted as an illustrative case to maximize the contrast between the two dark matter models. Both simulations adopt the same cosmological parameters, $\Omega_{\rm m}=0.25$, $\Omega_{\Lambda}=0.75$, $\sigma_{8}=0.9$, and $h=0.73$.

\subsection{21cm Brightness Temperature Modelling}

The hydrodynamical simulations provide the spatial distributions of gas density and temperature. However, the hydrogen ionization fractions are recalculated in post-processing due to absence of self-shielding against the UV background, which is significant to 21cm brightness temperature mock. The attenuation of the photoionization rate is modeled following the fitting formula from \citet{Rahmati2013}. The ionization equilibrium equation is then solved using the corrected photoionization rate, yielding updated neutral hydrogen (HI) fractions for all gas elements.

In the post-reionization era, the intergalactic gas is optically thin, and the spin temperature $T_{\rm s}$ is assumed to dominate over the background radiation field ($T_{\rm s}\gg T_{\rm CMB}$), which allows the integrated 21cm brightness temperature to be directly related to the HI column density through \citep{Meyer2017}
\begin{equation}
\frac{N_{\rm HI}}{\rm cm^{-2}}= \frac{32\pi k\nu_{\mathrm{HI}}}{3A_{\mathrm{HI}}hc^{2}} \int T_{\rm b}\,{\rm d}\nu_{\rm rest} = 3.85\times10^{20} \int \left(\frac{T_{\rm b}}{\rm K}\right) \left(\frac{{\rm d}\nu_{\rm rest}}{\rm MHz}\right),
\label{eq: 21cm modeling}
\end{equation}
where $k$ is the Boltzmann constant, $h$ is the Planck constant, $c$ is the speed of light, $\nu_{\rm HI}=1420.4\,\mathrm{MHz}$ is the rest-frame frequency of the HI 21cm transition, $A_{\rm HI}$ is the spontaneous emission rate, $T_{\rm b}$ is the brightness temperature in the rest frame of the source, and ${\rm d}\nu_{\rm rest}$ is the frequency interval measured in that frame as well. This relation allows the intrinsic integrated 21cm brightness temperature maps to be constructed directly from the projected HI column density.

\subsection{21cm Mock Generation for SKA-Low}

The intrinsic 21cm brightness temperature map calculated following the previous section corresponds to observations with infinite angular resolution. In practice, however, the brightness temperature maps measured by radio telescopes have to account for the convolution of the intrinsic signal with the telescope beam, which leads to the loss of small-scale information. To mimic realistic observations, we convolve the intrinsic integrated 21cm brightness temperature maps with a two-dimensional Gaussian point spread function (PSF). The standard deviation $\sigma$ of the function is related to the beam size $s$ through $s=2\sqrt{2\ln2}\,\sigma$. The beam size defines the angular resolution of the observation and indicates the FWHM of the Gaussian synthesized beam.

To investigate the impact of angular resolution on the detectability of cosmic filaments, the convolution is performed using a series of beam sizes corresponding to different angular resolutions. The adopted minimum beam sizes correspond to angular resolutions which can be carried out by SKA1-Low with maximum baseline about 70 km. By convolving the intrinsic maps with Gaussian beams of different FWHM values, we quantify how beam smearing affects the observable 21cm emission from cosmic filaments. For each beam, we measure the mean value and value range of integrated 21cm brightness temperature within the filament region and compare it with the expected sensitivity of SKA1-Low and SKA2-Low. The sensitivity of radio telescope reads \citep{Meyer2017},
\begin{equation}
\sigma_{T_{\rm b,int}} = 1.34 \times 10^{3} 
\frac{(1+z)^4}{\eta \sqrt{N(N-1)}} 
\left( \frac{\Omega_{\rm bm}}{\text{arcsec}^2} \right)^{-1}
\left( \frac{A_e / T_{\rm sys}}{\text{m}^2 \text{K}^{-1}} \right)^{-1}
\left( \frac{\Delta t}{\text{s}} \right)^{-1/2}
\left( \frac{\Delta \nu}{\text{Hz}} \right)^{1/2},
\label{eq: thermal noise}
\end{equation}
where $\sigma_{T_{\rm b,int}}$ is the rms noise of the integrated 21cm brightness temperature shown in Eq.~\ref{eq: 21cm modeling}, $z$ is the redshift, $\eta$ is the system efficiency, $N$ is the number of antenna stations, $\Omega_{\rm bm}$ is the solid angle of the beam, $A_{\rm e}/T_{\rm sys}$ is sensitivity metric of each station, $\Delta t$ is the total integration time, and $\Delta \nu$ is the observational frequency bandwidth.

\section{Results}
\label{sect:res}

\begin{figure*}
\centering
\includegraphics[width=0.75\columnwidth]{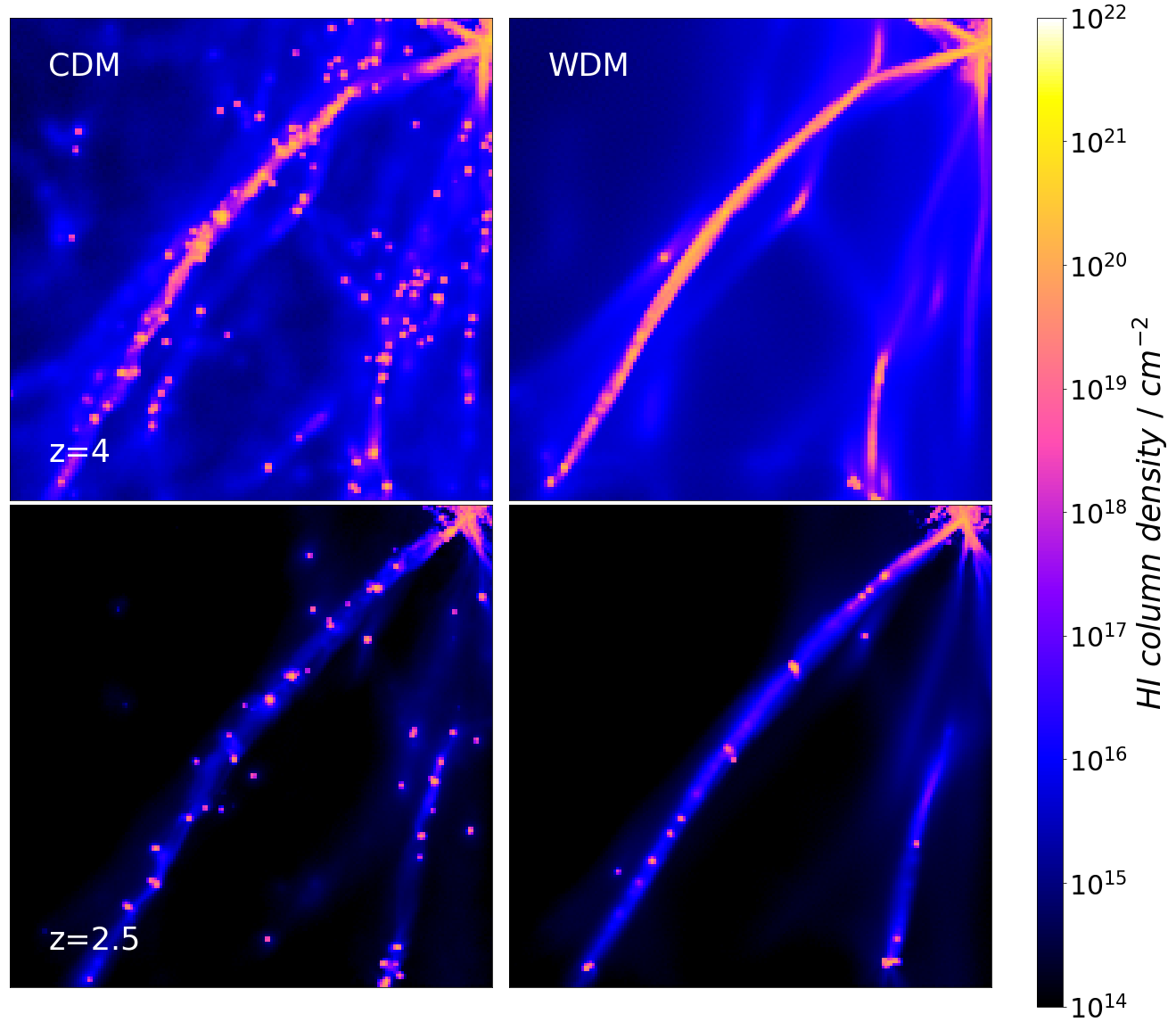}
\caption{Intrinsic HI column density of the same cosmic filament at $z=4$ and $z=2.5$ in the CDM (left) and WDM (right) zoom-in simulations. The HI filament of WDM appears smoother and contains fewer compact structures, especially at redshift $z=4$.}
\label{fig:HI_map}
\end{figure*}

\begin{figure*}
\centering
\includegraphics[width=0.75\columnwidth]{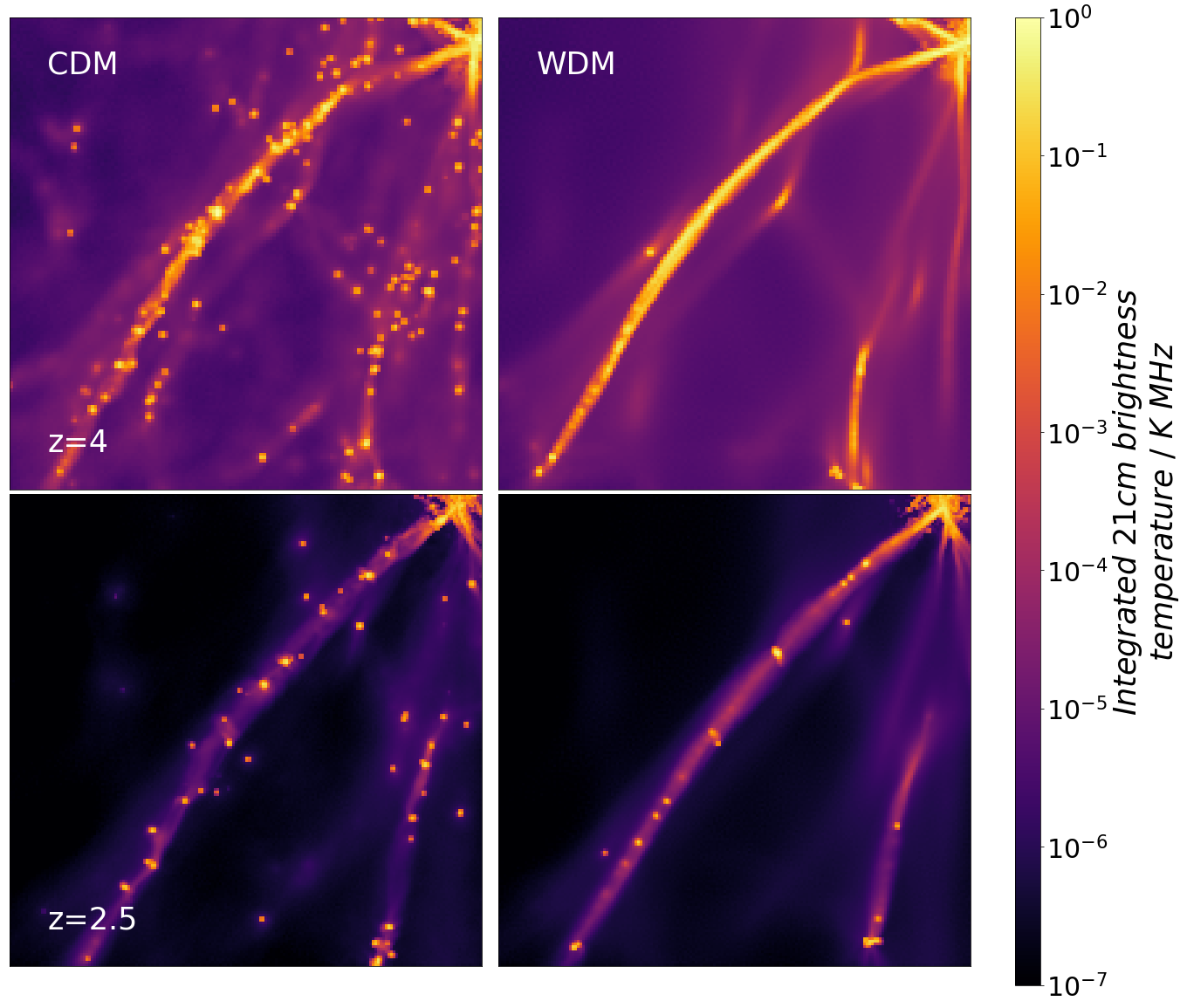}
\caption{Integrated 21cm brightness temperature of the same cosmic filament at $z=4$ and $z=2.5$ in the CDM (left) and WDM (right) zoom-in simulations. The morphology of 21cm filament closely follows the neutral hydrogen distribution.}
\label{fig:21cm_map}
\end{figure*}

\subsection{Filament Morphology}

Figs.~\ref{fig:HI_map} and~\ref{fig:21cm_map} present the intrinsic HI column density and the corresponding integrated 21cm brightness temperature maps of a representative filament at $z=4$ and $2.5$ for the CDM and WDM in zoom-in region. The maps cover a region of approximately $1.85\,h^{-1}\,\mathrm{cMpc}$ on a side, with a projection thickness of $2.5\,h^{-1}\,\mathrm{cMpc}$. The thickness was chosen to encompass the selected filament while reducing the frequency bandwidth and thermal noise. Since the thermal noise scales as $\sqrt{\Delta\nu}$, a narrower projection corresponds to a smaller frequency bandwidth and hence a lower thermal noise. In the CDM simulation, the filament exhibits a clumpy appearance, whereas the WDM filament is noticeably smoother and more continuous owing to the suppression of small-scale structure formation. The difference reduces toward lower redshift where some clumpy structures start to emerge in WDM case. The integrated 21cm brightness temperature closely follows the spatial distribution of neutral hydrogen due to its relative simple emission mechanism. Consequently, the differences visible in the HI column density are also reflected in the 21cm maps, suggesting that the morphology of filamentary 21cm emission retains information about the underlying dark matter model before instrumental effects are taken into account. Furthermore, the morphologies of 21cm filaments at high redshifts, e.g. $z=4$, are better suited as probes of dark matter properties. We therefore investigate the detectability of 21cm filaments at $z=4$ with SKA-Low.

\begin{figure*}
\centering
\includegraphics[width=\columnwidth]{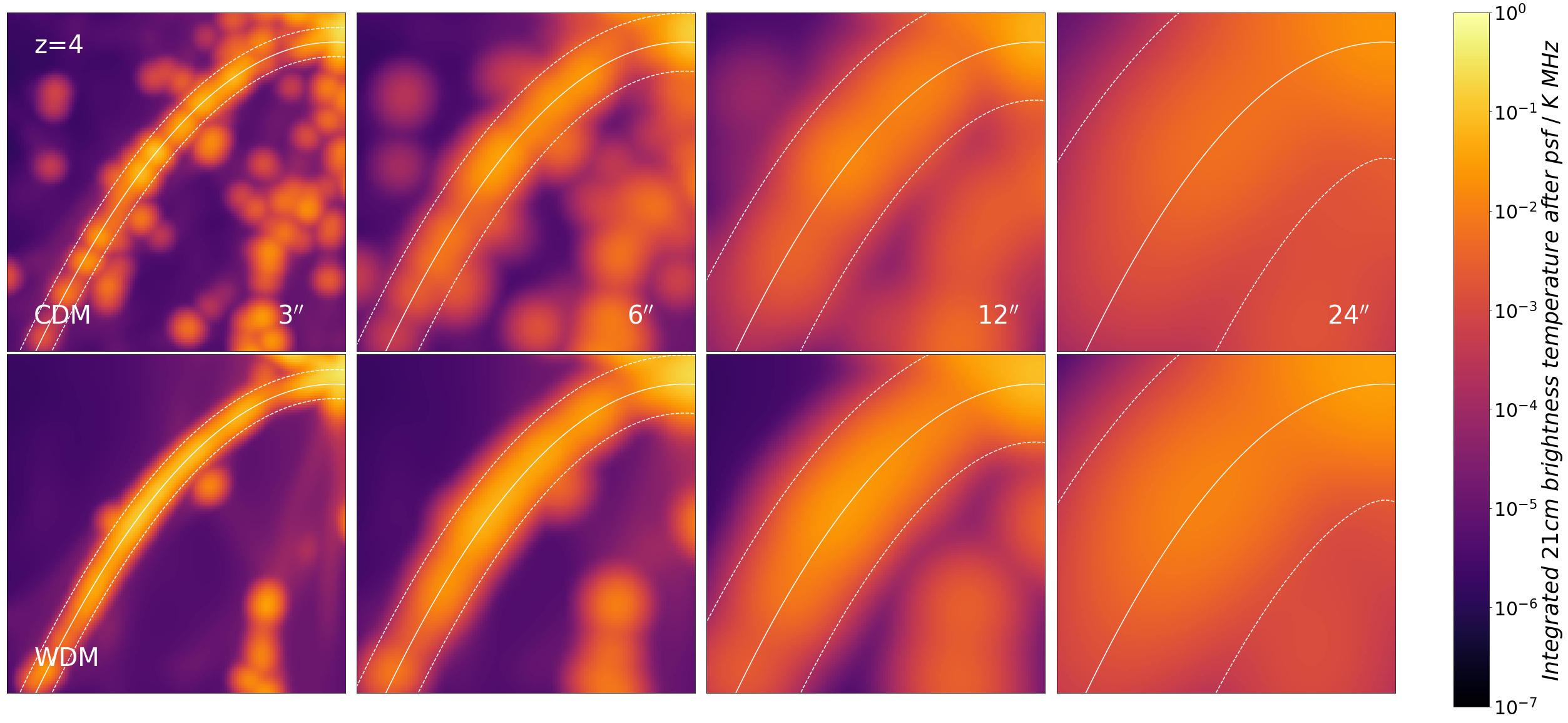}
\caption{Beam-convolved integrated 21cm brightness temperature maps of the filament shown in Figure~\ref{fig:21cm_map}. From left to right, the Gaussian beam size increases from 3\arcsec to 24\arcsec. The white solid curve trace the filament spine, while the white dashed curves mark the boundaries of the filament region. Beam smearing efficiently suppresses small-scale structures, particularly in the CDM case, reducing the observable morphological differences between the two dark matter models.}
\label{fig:21cm_psf}
\end{figure*}

\subsection{Beam Convolution}

The observed differences between the two dark matter models are expected to depend on the angular resolution. To quantify this effect, the intrinsic 21cm brightness temperature maps are convolved with Gaussian beams of increasing FWHM. The resulting mock observations of $z=4$ are shown in Fig.~\ref{fig:21cm_psf}. The smallest beam size adopted in this work corresponds to the highest angular resolution achievable by SKA1-Low at the observed wavelength of the redshifted 21cm emission. The filament in the figure is identified using the method of \citet{Liao2019}. Following \citet{Cautun2014}, all connected pixels belonging to the filament are compressed onto its central axis, which is subsequently fitted with a quadratic curve to define the filament spine (solid curve in Fig.~\ref{fig:21cm_psf}). The dashed curves mark the filament region used for the brightness measurements, with a width equal to two times FWHM of the adopted Gaussian beam.

As the beam size increases, small-scale structures are progressively smoothed out in both dark matter models. The compact bright clumps visible in the intrinsic CDM map rapidly disappear and become almost indistinguishable from the WDM case for beam sizes of $6\arcsec$ and larger. Although the filament itself remains identifiable after beam convolution, its internal structure is largely erased, substantially reducing the morphological differences between the two models. To further assess the observational prospects, we measure the integrated 21cm brightness temperature within the filament region for each beam size. This allows us to determine if we can detect this structure with high resolution.

\begin{figure*}
\centering
\includegraphics[width=0.75\columnwidth]{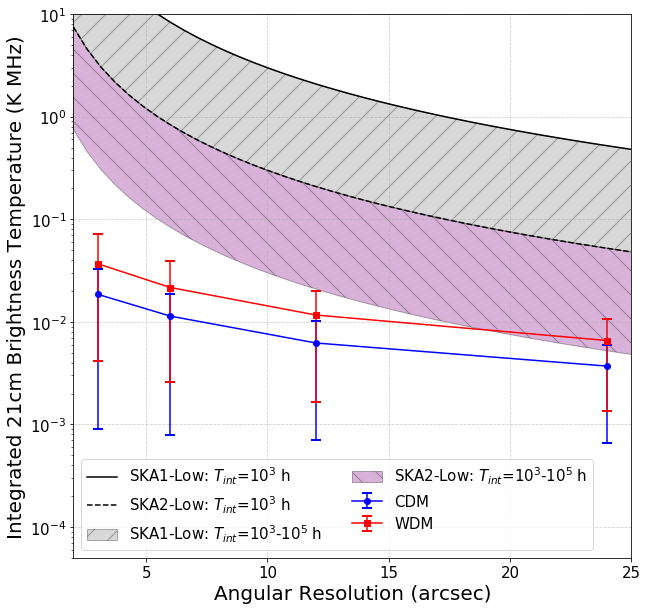}
\caption{Integrated 21cm brightness temperature measured within the filament region as a function of beam size for the CDM (blue) and WDM (red) simulations. Symbols show the mean brightness temperature, and error bars indicate the 20th$-$80th percentile range measured within the filament region. The shaded regions show the expected sensitivities of SKA1-Low (gray) and SKA2-Low (purple) for different integration time ($T_{\rm int}$) and beam size. Although the WDM filament remains systematically brighter than the CDM filament at all angular resolutions, neither signal exceeds the expected sensitivity of SKA1-Low or SKA2-Low at the angular resolutions required to preserve the filamentary morphology. Therefore, direct high-angular-resolution imaging of individual filaments at $z=4$ is expected to remain extremely challenging, even with next-generation radio facilities.}
\label{fig:21cm_ska}
\end{figure*}

\begin{table}
\begin{center}
\caption[]{The adopted parameters of SKA1-Low.}
\label{tab:para_ska}
 \begin{tabular}{cccc}
  \hline\noalign{\smallskip}
Telescope &  Maximum Baseline [km]  & Number of Station & Sensitivity $A_{\rm e}/T_{\rm sys}$ [$\rm m^{2}/K$]    \\
\hline\noalign{\smallskip}
SKA1-Low  & 70 &  512  &  1.3092  \\ 
\noalign{\smallskip}\hline
\end{tabular}
\end{center}
\end{table}

\subsection{SKA Detectability}

Fig.~\ref{fig:21cm_ska} compares the integrated 21cm brightness temperature measured along the filament with the expected sensitivity of SKA1-Low and SKA2-Low for different angular resolutions. The shaded regions show the expected sensitivities of SKA1-Low (gray) and SKA2-Low (purple) for different integration times and beam sizes. The adopted parameters for SKA1-Low are summarized in Tab.~\ref{tab:para_ska} and are taken from \citet{Braun2019}. The single-station sensitivity metric is interpolated from Table~9 at the observational frequency of $\nu_{\rm obs}=284\,\mathrm{MHz}$. We adopt $\eta=1$ for an idealized sensitivity estimate, and the bandwidth is set to $\Delta\nu=0.268\,\mathrm{MHz}$, corresponding to the line-of-sight thickness of the projected map. For SKA2-Low, we assume an approximate tenfold improvement in sensitivity relative to SKA1-Low, based on its planned design of approximately 4880 low-frequency stations \citep{Braun2019}. For both dark matter models, the measured brightness decreases monotonically with increasing beam size. This behavior is expected because beam convolution averages the filament emission over a larger solid angle, reducing the observed surface brightness. At all angular resolutions, the WDM filament remains systematically brighter than its CDM counterpart.

Also shown are the SKA-Low sensitivities for different observing times. Although the intrinsic filament signal is strongest at the highest angular resolution, the corresponding instrumental sensitivity is insufficient for a significant detection, especially when resolution higher than 12\arcsec. Conversely, at larger beam sizes where the telescope sensitivity improves and the signal from filament starts to be detected for SKA2-Low, where beam smearing substantially suppresses the filament signal. However, the filament emission remains below the SKA1-Low sensitivity over the full range of angular resolutions considered in this work. These results indicate that, despite the intrinsic differences between CDM and WDM filamentary 21cm emission, directly resolving individual filaments at $z=4$ with current SKA1-Low capabilities is expected to be impossible, even with SKA2-Low it is still extremely challenging.

We note that our sensitivity estimates represent optimistic limits. We adopt an idealized system efficiency of $\eta=1$, while more realistic values of $\eta\sim0.5$--$0.8$ would increase the noise by a factor of $\sim1.25$--$2$. In addition, only thermal noise is considered, while explicitly model Galactic and extragalactic foregrounds, foreground residuals and other instrumental systematics are not included. These effects would further make the direct detection of individual 21cm filaments even more challenging.

Finally, we note that the $1.5\,\mathrm{keV}$ WDM model adopted here lies below current Ly$\alpha$ forest constraints \citep{Villasenor2023, Irsic2024} and is used as an illustrative case to maximize the contrast with CDM. For more realistic WDM masses of $\sim3$--$5\,\mathrm{keV}$, the suppression of small-scale structure and the resulting differences in filament morphology are expected to be weaker. Since direct imaging is already extremely challenging in the favorable $1.5\,\mathrm{keV}$ case, distinguishing more realistic WDM models through the morphology of individual filaments is expected to be even more difficult.

\section{Discussion and Conclusion}
\label{sect:discussion}

In this work, we investigate the feasibility of probing dark matter models through the morphology of filamentary 21cm emission at $z=4$. Using cosmological hydrodynamical simulations of the CDM and WDM scenarios, we construct intrinsic HI column density and integrated 21cm brightness temperature maps, and further generate mock observations by convolving the intrinsic emission with Gaussian beams corresponding to different angular resolutions. The resulting filament brightness is compared with the expected sensitivities of SKA1-Low and SKA2-Low. We find that WDM filaments are intrinsically smoother and systematically brighter than their CDM counterparts. However, these morphological differences rapidly disappear as the beam size increases. 
Although SKA2-Low can detect the filament with low angular resolution, distinguishing the two dark matter models requires a resolution so high that even SKA2-Low would fail to detect the signal.
Consequently, direct high-angular-resolution imaging of individual filaments is unlikely to provide useful constraints on dark matter with current and planned SKA-Low facilities.

Several sources of uncertainty should be considered when interpreting these results. First, the simulations do not include stellar or AGN feedback, both of which can modify the gas distribution and ionization state within and around cosmic filaments. Radiation from galaxies and AGN may further alter the neutral hydrogen fraction, introducing additional uncertainties in the predicted 21cm brightness temperature, although they are expected to have a smaller impact on the large-scale morphological differences investigated in this work. Moreover, the present simulations adopt a simplified treatment of star formation and neglect emission from star-forming gas. This physical process may enhance the absolute brightness of the filamentary 21cm emission, reducing the difficulty of detecting the filamentary structures. Nevertheless, our conclusions regarding the difficulty of directly imaging filamentary 21cm emission are unlikely to change qualitatively.

Our analysis is based on a single representative filament and therefore does not establish the statistical universality of the morphological differences between CDM and WDM. A larger sample may yield a wider range of filament brightness and morphology, including brighter structures. Nevertheless, our results indicate that even for this relatively favorable case, resolving individual filaments while achieving sufficient 21cm surface-brightness sensitivity is challenging with SKA-Low, suggesting that this observational difficulty is unlikely to be altered by sample variance. 

Although direct high-resolution imaging of individual filaments appears extremely challenging, filamentary 21cm emission remains a promising probe of the cosmic web and dark matter physics. Future radio facilities with greatly improved sensitivity and angular resolution, combined with other observational strategies such as statistical stacking \citep{Meng2026}, may significantly enhance the detectability of diffuse filamentary structures. Alongside mainstream 21cm probes like the power spectrum \citep{Carucci2015, Bauer2021}, these advancements will enable further exploration of the nature of dark matter.

\begin{acknowledgements}
We thank Liang Gao and Shihong Liao for helpful discussions and valuable comments on the manuscript. We acknowledge the supports from the National Natural Science Foundation of China (Grant No. 12588202) and the National Key Research and Development Program of China (Grant No. 2023YFB3002500). H.H. is supported by the NSFC Grant Nos. 12503012.
\end{acknowledgements}

\bibliographystyle{aasjournal}
\bibliography{bibtex}

\begin{thebibliography}{}
\expandafter\ifx\csname natexlab\endcsname\relax\def\natexlab#1{#1}\fi
\providecommand{\url}[1]{\href{#1}{#1}}
\providecommand{\dodoi}[1]{doi:~\href{http://doi.org/#1}{\nolinkurl{#1}}}
\providecommand{\doeprint}[1]{\href{http://ascl.net/#1}{\nolinkurl{http://ascl.net/#1}}}
\providecommand{\doarXiv}[1]{\href{https://arxiv.org/abs/#1}{\nolinkurl{https://arxiv.org/abs/#1}}}

\bibitem[{{Avila-Reese} {et~al.}(2001){Avila-Reese}, {Col{\'\i}n}, {Valenzuela}, {D'Onghia}, \& {Firmani}}]{Avila-Reese2001}
{Avila-Reese}, V., {Col{\'\i}n}, P., {Valenzuela}, O., {D'Onghia}, E., \& {Firmani}, C. 2001, \apj, 559, 516, \dodoi{10.1086/322411}

\bibitem[{{Bauer} {et~al.}(2021){Bauer}, {Marsh}, {Hlo{\v{z}}ek}, {Padmanabhan}, \& {Lagu{\"e}}}]{Bauer2021}
{Bauer}, J.~B., {Marsh}, D. J.~E., {Hlo{\v{z}}ek}, R., {Padmanabhan}, H., \& {Lagu{\"e}}, A. 2021, \mnras, 500, 3162, \dodoi{10.1093/mnras/staa3300}

\bibitem[{{Baur} {et~al.}(2016){Baur}, {Palanque-Delabrouille}, {Y{\`e}che}, {Magneville}, \& {Viel}}]{Baur2016}
{Baur}, J., {Palanque-Delabrouille}, N., {Y{\`e}che}, C., {Magneville}, C., \& {Viel}, M. 2016, \jcap, 2016, 012, \dodoi{10.1088/1475-7516/2016/08/012}

\bibitem[{{Bode} {et~al.}(2001){Bode}, {Ostriker}, \& {Turok}}]{Bode2001}
{Bode}, P., {Ostriker}, J.~P., \& {Turok}, N. 2001, \apj, 556, 93, \dodoi{10.1086/321541}

\bibitem[{{Bond} {et~al.}(1996){Bond}, {Kofman}, \& {Pogosyan}}]{Bond1996}
{Bond}, J.~R., {Kofman}, L., \& {Pogosyan}, D. 1996, \nat, 380, 603, \dodoi{10.1038/380603a0}

\bibitem[{{Bose} {et~al.}(2017){Bose}, {Hellwing}, {Frenk}, {Jenkins}, {Lovell}, {Helly}, {Li}, {Gonzalez-Perez}, \& {Gao}}]{Bose2017}
{Bose}, S., {Hellwing}, W.~A., {Frenk}, C.~S., {et~al.} 2017, \mnras, 464, 4520, \dodoi{10.1093/mnras/stw2686}

\bibitem[{{Boyarsky} {et~al.}(2009){Boyarsky}, {Lesgourgues}, {Ruchayskiy}, \& {Viel}}]{Boyarsky2009}
{Boyarsky}, A., {Lesgourgues}, J., {Ruchayskiy}, O., \& {Viel}, M. 2009, \jcap, 2009, 012, \dodoi{10.1088/1475-7516/2009/05/012}

\bibitem[{{Braun} {et~al.}(2019){Braun}, {Bonaldi}, {Bourke}, {Keane}, \& {Wagg}}]{Braun2019}
{Braun}, R., {Bonaldi}, A., {Bourke}, T., {Keane}, E., \& {Wagg}, J. 2019, arXiv e-prints, arXiv:1912.12699, \dodoi{10.48550/arXiv.1912.12699}

\bibitem[{{Byrohl} \& {Nelson}(2023)}]{Byrohl2023}
{Byrohl}, C., \& {Nelson}, D. 2023, \mnras, 523, 5248, \dodoi{10.1093/mnras/stad1779}

\bibitem[{{Carucci} {et~al.}(2015){Carucci}, {Villaescusa-Navarro}, {Viel}, \& {Lapi}}]{Carucci2015}
{Carucci}, I.~P., {Villaescusa-Navarro}, F., {Viel}, M., \& {Lapi}, A. 2015, \jcap, 2015, 047, \dodoi{10.1088/1475-7516/2015/07/047}

\bibitem[{Cautun(2014)}]{Cautun2014}
Cautun, M. 2014, PhD thesis, University of Groningen

\bibitem[{{Col{\'\i}n} {et~al.}(2000){Col{\'\i}n}, {Avila-Reese}, \& {Valenzuela}}]{Colin2000}
{Col{\'\i}n}, P., {Avila-Reese}, V., \& {Valenzuela}, O. 2000, \apj, 542, 622, \dodoi{10.1086/317057}

\bibitem[{{Diemand} {et~al.}(2005){Diemand}, {Moore}, \& {Stadel}}]{Diemand2005}
{Diemand}, J., {Moore}, B., \& {Stadel}, J. 2005, \nat, 433, 389, \dodoi{10.1038/nature03270}

\bibitem[{{Elias} {et~al.}(2020){Elias}, {Genel}, {Sternberg}, {Devriendt}, {Slyz}, {Visbal}, \& {Bouch{\'e}}}]{Elias2020}
{Elias}, L.~M., {Genel}, S., {Sternberg}, A., {et~al.} 2020, \mnras, 494, 5439, \dodoi{10.1093/mnras/staa1059}

\bibitem[{{Frenk} \& {White}(2012)}]{Frenk2012}
{Frenk}, C.~S., \& {White}, S.~D.~M. 2012, Annalen der Physik, 524, 507, \dodoi{10.1002/andp.201200212}

\bibitem[{{Gao} \& {Theuns}(2007)}]{Gao2007}
{Gao}, L., \& {Theuns}, T. 2007, Science, 317, 1527, \dodoi{10.1126/science.1146676}

\bibitem[{{Gao} {et~al.}(2015){Gao}, {Theuns}, \& {Springel}}]{Gao2015}
{Gao}, L., {Theuns}, T., \& {Springel}, V. 2015, \mnras, 450, 45, \dodoi{10.1093/mnras/stv643}

\bibitem[{{Garzilli} {et~al.}(2017){Garzilli}, {Boyarsky}, \& {Ruchayskiy}}]{Garzilli2017}
{Garzilli}, A., {Boyarsky}, A., \& {Ruchayskiy}, O. 2017, Physics Letters B, 773, 258, \dodoi{10.1016/j.physletb.2017.08.022}

\bibitem[{{Garzilli} {et~al.}(2021){Garzilli}, {Magalich}, {Ruchayskiy}, \& {Boyarsky}}]{Garzilli2021}
{Garzilli}, A., {Magalich}, A., {Ruchayskiy}, O., \& {Boyarsky}, A. 2021, \mnras, 502, 2356, \dodoi{10.1093/mnras/stab192}

\bibitem[{{Garzilli} {et~al.}(2019){Garzilli}, {Magalich}, {Theuns}, {Frenk}, {Weniger}, {Ruchayskiy}, \& {Boyarsky}}]{Garzilli2019}
{Garzilli}, A., {Magalich}, A., {Theuns}, T., {et~al.} 2019, \mnras, 489, 3456, \dodoi{10.1093/mnras/stz2188}

\bibitem[{{Gilman} {et~al.}(2020){Gilman}, {Birrer}, {Nierenberg}, {Treu}, {Du}, \& {Benson}}]{Gilman2020}
{Gilman}, D., {Birrer}, S., {Nierenberg}, A., {et~al.} 2020, \mnras, 491, 6077, \dodoi{10.1093/mnras/stz3480}

\bibitem[{{Green} {et~al.}(2004){Green}, {Hofmann}, \& {Schwarz}}]{Green2004}
{Green}, A.~M., {Hofmann}, S., \& {Schwarz}, D.~J. 2004, \mnras, 353, L23, \dodoi{10.1111/j.1365-2966.2004.08232.x}

\bibitem[{{Haardt} \& {Madau}(2001)}]{Haardt2001}
{Haardt}, F., \& {Madau}, P. 2001, in Clusters of Galaxies and the High Redshift Universe Observed in X-rays, ed. D.~M. {Neumann} \& J.~T.~V. {Tran}, 64, \dodoi{10.48550/arXiv.astro-ph/0106018}

\bibitem[{{Hernquist} {et~al.}(1996){Hernquist}, {Katz}, {Weinberg}, \& {Miralda-Escud{\'e}}}]{Hernquist1996}
{Hernquist}, L., {Katz}, N., {Weinberg}, D.~H., \& {Miralda-Escud{\'e}}, J. 1996, \apjl, 457, L51, \dodoi{10.1086/309899}

\bibitem[{{Hezaveh} {et~al.}(2016{\natexlab{a}}){Hezaveh}, {Dalal}, {Holder}, {Kisner}, {Kuhlen}, \& {Perreault Levasseur}}]{Hezaveh2016_b}
{Hezaveh}, Y., {Dalal}, N., {Holder}, G., {et~al.} 2016{\natexlab{a}}, \jcap, 2016, 048, \dodoi{10.1088/1475-7516/2016/11/048}

\bibitem[{{Hezaveh} {et~al.}(2016{\natexlab{b}}){Hezaveh}, {Dalal}, {Marrone}, {Mao}, {Morningstar}, {Wen}, {Blandford}, {Carlstrom}, {Fassnacht}, {Holder}, {Kemball}, {Marshall}, {Murray}, {Perreault Levasseur}, {Vieira}, \& {Wechsler}}]{Hezaveh2016_a}
{Hezaveh}, Y.~D., {Dalal}, N., {Marrone}, D.~P., {et~al.} 2016{\natexlab{b}}, \apj, 823, 37, \dodoi{10.3847/0004-637X/823/1/37}

\bibitem[{{Hofmann} {et~al.}(2001){Hofmann}, {Schwarz}, \& {St{\"o}cker}}]{Hofmann2001}
{Hofmann}, S., {Schwarz}, D.~J., \& {St{\"o}cker}, H. 2001, \prd, 64, 083507, \dodoi{10.1103/PhysRevD.64.083507}

\bibitem[{{Ir{\v{s}}i{\v{c}}} {et~al.}(2017){Ir{\v{s}}i{\v{c}}}, {Viel}, {Haehnelt}, {Bolton}, {Cristiani}, {Becker}, {D'Odorico}, {Cupani}, {Kim}, {Berg}, {L{\'o}pez}, {Ellison}, {Christensen}, {Denney}, \& {Worseck}}]{Irsic2017}
{Ir{\v{s}}i{\v{c}}}, V., {Viel}, M., {Haehnelt}, M.~G., {et~al.} 2017, \prd, 96, 023522, \dodoi{10.1103/PhysRevD.96.023522}

\bibitem[{{Ir{\v{s}}i{\v{c}}} {et~al.}(2024){Ir{\v{s}}i{\v{c}}}, {Viel}, {Haehnelt}, {Bolton}, {Molaro}, {Puchwein}, {Boera}, {Becker}, {Gaikwad}, {Keating}, \& {Kulkarni}}]{Irsic2024}
---. 2024, \prd, 109, 043511, \dodoi{10.1103/PhysRevD.109.043511}

\bibitem[{{Khimey} {et~al.}(2021){Khimey}, {Bose}, \& {Tacchella}}]{Khimey2021}
{Khimey}, D., {Bose}, S., \& {Tacchella}, S. 2021, \mnras, 506, 4139, \dodoi{10.1093/mnras/stab2019}

\bibitem[{{Li} {et~al.}(2016){Li}, {Frenk}, {Cole}, {Gao}, {Bose}, \& {Hellwing}}]{Li2016}
{Li}, R., {Frenk}, C.~S., {Cole}, S., {et~al.} 2016, \mnras, 460, 363, \dodoi{10.1093/mnras/stw939}

\bibitem[{{Liao} \& {Gao}(2019)}]{Liao2019}
{Liao}, S., \& {Gao}, L. 2019, \mnras, 485, 464, \dodoi{10.1093/mnras/stz441}

\bibitem[{{Liu} {et~al.}(2025){Liu}, {Gao}, {Liao}, \& {Zhu}}]{Liu2025}
{Liu}, Y., {Gao}, L., {Liao}, S., \& {Zhu}, K. 2025, \apj, 984, 55, \dodoi{10.3847/1538-4357/adc44b}

\bibitem[{{Liu} {et~al.}(2026){Liu}, {Gao}, {Liao}, {Zhu}, {Jing}, \& {Hu}}]{Liu2026}
{Liu}, Y., {Gao}, L., {Liao}, S., {et~al.} 2026, arXiv e-prints, arXiv:2601.22677, \dodoi{10.48550/arXiv.2601.22677}

\bibitem[{{Liu} {et~al.}(2024){Liu}, {Gao}, {Bose}, {Frenk}, {Jenkins}, {Springel}, {Wang}, {White}, \& {Zheng}}]{Liu2024}
{Liu}, Y., {Gao}, L., {Bose}, S., {et~al.} 2024, \mnras, 527, 11740, \dodoi{10.1093/mnras/stae003}

\bibitem[{{Lovell} {et~al.}(2014){Lovell}, {Frenk}, {Eke}, {Jenkins}, {Gao}, \& {Theuns}}]{Lovell2014}
{Lovell}, M.~R., {Frenk}, C.~S., {Eke}, V.~R., {et~al.} 2014, \mnras, 439, 300, \dodoi{10.1093/mnras/stt2431}

\bibitem[{{Lovell} {et~al.}(2017){Lovell}, {Gonzalez-Perez}, {Bose}, {Boyarsky}, {Cole}, {Frenk}, \& {Ruchayskiy}}]{Lovell2017}
{Lovell}, M.~R., {Gonzalez-Perez}, V., {Bose}, S., {et~al.} 2017, \mnras, 468, 2836, \dodoi{10.1093/mnras/stx621}

\bibitem[{{Ludlow} {et~al.}(2016){Ludlow}, {Bose}, {Angulo}, {Wang}, {Hellwing}, {Navarro}, {Cole}, \& {Frenk}}]{Ludlow2016}
{Ludlow}, A.~D., {Bose}, S., {Angulo}, R.~E., {et~al.} 2016, \mnras, 460, 1214, \dodoi{10.1093/mnras/stw1046}

\bibitem[{{Meng} {et~al.}(2026){Meng}, {Wang}, {Jing}, {Chen}, \& {Liu}}]{Meng2026}
{Meng}, Y., {Wang}, J., {Jing}, Y., {Chen}, H., \& {Liu}, Z. 2026, \apj, 1002, 210, \dodoi{10.3847/1538-4357/ae61a5}

\bibitem[{{Metcalf} \& {Madau}(2001)}]{Metcalf2001}
{Metcalf}, R.~B., \& {Madau}, P. 2001, \apj, 563, 9, \dodoi{10.1086/323695}

\bibitem[{{Meyer} {et~al.}(2017){Meyer}, {Robotham}, {Obreschkow}, {Westmeier}, {Duffy}, \& {Staveley-Smith}}]{Meyer2017}
{Meyer}, M., {Robotham}, A., {Obreschkow}, D., {et~al.} 2017, \pasa, 34, 52, \dodoi{10.1017/pasa.2017.31}

\bibitem[{{Minor} {et~al.}(2017){Minor}, {Kaplinghat}, \& {Li}}]{Minor2017}
{Minor}, Q.~E., {Kaplinghat}, M., \& {Li}, N. 2017, \apj, 845, 118, \dodoi{10.3847/1538-4357/aa7fee}

\bibitem[{{Miranda} \& {Macci{\`o}}(2007)}]{Miranda2007}
{Miranda}, M., \& {Macci{\`o}}, A.~V. 2007, \mnras, 382, 1225, \dodoi{10.1111/j.1365-2966.2007.12440.x}

\bibitem[{{Mocz} {et~al.}(2019){Mocz}, {Fialkov}, {Vogelsberger}, {Becerra}, {Amin}, {Bose}, {Boylan-Kolchin}, {Chavanis}, {Hernquist}, {Lancaster}, {Marinacci}, {Robles}, \& {Zavala}}]{Mocz2019}
{Mocz}, P., {Fialkov}, A., {Vogelsberger}, M., {et~al.} 2019, \prl, 123, 141301, \dodoi{10.1103/PhysRevLett.123.141301}

\bibitem[{{Paduroiu}(2022)}]{Paduroiu2022}
{Paduroiu}, S. 2022, Universe, 8, 76, \dodoi{10.3390/universe8020076}

\bibitem[{{Rahmati} {et~al.}(2013){Rahmati}, {Pawlik}, {Rai{\v{c}}evi{\'c}}, \& {Schaye}}]{Rahmati2013}
{Rahmati}, A., {Pawlik}, A.~H., {Rai{\v{c}}evi{\'c}}, M., \& {Schaye}, J. 2013, \mnras, 430, 2427, \dodoi{10.1093/mnras/stt066}

\bibitem[{{Schneider} {et~al.}(2012){Schneider}, {Smith}, {Macci{\`o}}, \& {Moore}}]{Schneider2012}
{Schneider}, A., {Smith}, R.~E., {Macci{\`o}}, A.~V., \& {Moore}, B. 2012, \mnras, 424, 684, \dodoi{10.1111/j.1365-2966.2012.21252.x}

\bibitem[{{Springel}(2005)}]{Springel2005}
{Springel}, V. 2005, \mnras, 364, 1105, \dodoi{10.1111/j.1365-2966.2005.09655.x}

\bibitem[{{Springel} {et~al.}(2006){Springel}, {Frenk}, \& {White}}]{Springel2006}
{Springel}, V., {Frenk}, C.~S., \& {White}, S. D.~M. 2006, \nat, 440, 1137, \dodoi{10.1038/nature04805}

\bibitem[{{Springel} {et~al.}(2008){Springel}, {Wang}, {Vogelsberger}, {Ludlow}, {Jenkins}, {Helmi}, {Navarro}, {Frenk}, \& {White}}]{Springel2008}
{Springel}, V., {Wang}, J., {Vogelsberger}, M., {et~al.} 2008, \mnras, 391, 1685, \dodoi{10.1111/j.1365-2966.2008.14066.x}

\bibitem[{{Vegetti} {et~al.}(2012){Vegetti}, {Lagattuta}, {McKean}, {Auger}, {Fassnacht}, \& {Koopmans}}]{Vegetti2012}
{Vegetti}, S., {Lagattuta}, D.~J., {McKean}, J.~P., {et~al.} 2012, \nat, 481, 341, \dodoi{10.1038/nature10669}

\bibitem[{{Viel} {et~al.}(2013){Viel}, {Becker}, {Bolton}, \& {Haehnelt}}]{Viel2013}
{Viel}, M., {Becker}, G.~D., {Bolton}, J.~S., \& {Haehnelt}, M.~G. 2013, \prd, 88, 043502, \dodoi{10.1103/PhysRevD.88.043502}

\bibitem[{{Viel} {et~al.}(2004){Viel}, {Haehnelt}, \& {Springel}}]{Viel2004}
{Viel}, M., {Haehnelt}, M.~G., \& {Springel}, V. 2004, \mnras, 354, 684, \dodoi{10.1111/j.1365-2966.2004.08224.x}

\bibitem[{{Viel} {et~al.}(2005){Viel}, {Lesgourgues}, {Haehnelt}, {Matarrese}, \& {Riotto}}]{Viel2005}
{Viel}, M., {Lesgourgues}, J., {Haehnelt}, M.~G., {Matarrese}, S., \& {Riotto}, A. 2005, \prd, 71, 063534, \dodoi{10.1103/PhysRevD.71.063534}

\bibitem[{{Villasenor} {et~al.}(2023){Villasenor}, {Robertson}, {Madau}, \& {Schneider}}]{Villasenor2023}
{Villasenor}, B., {Robertson}, B., {Madau}, P., \& {Schneider}, E. 2023, \prd, 108, 023502, \dodoi{10.1103/PhysRevD.108.023502}

\bibitem[{{Wang} {et~al.}(2020){Wang}, {Bose}, {Frenk}, {Gao}, {Jenkins}, {Springel}, \& {White}}]{Wang2020}
{Wang}, J., {Bose}, S., {Frenk}, C.~S., {et~al.} 2020, \nat, 585, 39, \dodoi{10.1038/s41586-020-2642-9}

\bibitem[{{Witstok} {et~al.}(2021){Witstok}, {Puchwein}, {Kulkarni}, {Smit}, \& {Haehnelt}}]{Witstok2021}
{Witstok}, J., {Puchwein}, E., {Kulkarni}, G., {Smit}, R., \& {Haehnelt}, M.~G. 2021, \aap, 650, A98, \dodoi{10.1051/0004-6361/202040187}

\bibitem[{{Zackrisson} \& {Riehm}(2010)}]{Zackrisson2010}
{Zackrisson}, E., \& {Riehm}, T. 2010, Advances in Astronomy, 2010, 478910, \dodoi{10.1155/2010/478910}

\bibitem[{{Zheng} {et~al.}(2024){Zheng}, {Bose}, {Frenk}, {Gao}, {Jenkins}, {Liao}, {Liu}, \& {Wang}}]{Zheng2024}
{Zheng}, H., {Bose}, S., {Frenk}, C.~S., {et~al.} 2024, \mnras, 528, 7300, \dodoi{10.1093/mnras/stae289}

\end{thebibliography}

\label{lastpage}

\end{document}